\documentclass[11pt]{article}
\usepackage{epsfig}
\newcommand{\sss}{\vspace{.2in}}

\newcommand{\be}{\begin{equation}}
\newcommand{\ee}{\end{equation}}
\newcommand{\bea}{\begin{eqnarray}}
\newcommand{\eea}{\end{eqnarray}}
\newcommand{\sn}{{\rm sn}}
\newcommand{\cn}{{\rm cn}}
\newcommand{\dn}{{\rm dn}}
\newcommand{\sech}{{\rm sech}}

\begin{document}
\vspace{.2in}
~\hfill{\footnotesize }
\vspace{.5in}
\begin{center}
{\LARGE {\bf A Generalization of Landen's Quadratic Transformation Formulas for Jacobi Elliptic Functions}}
\end{center}
\vspace{.5in}
\begin{center}
{\large{\bf
   \mbox{Avinash Khare}$^{a,}$\footnote{khare@iopb.res.in} and
   \mbox{Uday Sukhatme}$^{b,}$\footnote{sukhatme@uic.edu}
 }}
\end{center}
\vspace{.6in}
\noindent
a) \hspace*{.2in}
Institute of Physics, Sachivalaya Marg, Bhubaneswar 751005, Orissa, India\\
b) \hspace*{.2in}
Department of Physics, University of Illinois at Chicago, Chicago, IL 60607-7059, U.S.A. \\
\sss
\sss
\vspace{1.1in}
\begin{center}
{\Large {\bf Abstract}}
\end{center}
Landen formulas, which connect Jacobi elliptic functions with different 
modulus parameters, were first obtained over two hundred years ago by making 
a suitable quadratic transformation of variables in elliptic integrals. We 
obtain and discuss significant generalizations of the celebrated Landen 
formulas. Our approach is based on some recently obtained periodic solutions 
of physically interesting nonlinear differential equations and numerous 
remarkable new cyclic identities involving Jacobi elliptic functions. 
\newpage

\sss

Jacobi elliptic functions $\sn(x,m)$, $\cn(x,m)$ and $\dn(x,m)$, with 
elliptic modulus parameter $m \equiv k^2~(0 \le m \le 1)$ play an important role in describing periodic solutions of many linear and nonlinear differential equations of interest in diverse branches of engineering, physics and mathematics. The Jacobi elliptic functions are often defined 
with the help of the elliptic integral
$$ 
\int \frac{dz}{\sqrt{(1-z^2)(1-k^2z^2)}}~~. 
$$ 
Over two centuries ago, John Landen \cite{lan} 
studied the quadratic transformation of 
variables 
$$ 
t = \frac{(1+k')~z~\sqrt{1-z^2}}{\sqrt{1-k^2z^2}}~~,
~k' \equiv \sqrt{1-k^2}=\sqrt{1-m}~~. 
$$ 
This transformation yields the transformed integral 
$$ 
\int \frac{dt}{(1+k') \sqrt{(1-t^2)(1-l^2t^2)}}~~, 
~~l \equiv \frac{1-k'}{1+k'}~~. 
$$ 
It readily follows that \cite{han}
\be\label{E1} 
\sn\left[(1+k')u, \left(\frac{1-k'}{1+k'}\right)^2\right]=
\frac{(1+k')~\sn(u,m)~\cn(u,m)}{\dn(u,m)}~,\\
\ee
\be\label{E2}
\cn\left[(1+k')u, \left(\frac{1-k'}{1+k'}\right)^2\right]
=\frac{1-(1+k')~\sn^2(u,m)}{\dn(u,m)}~,\\
\ee
\be\label{E3}
\dn\left[(1+k')u, \left(\frac{1-k'}{1+k'}\right)^2\right]
=\frac{1-(1-k')~\sn^2(u,m)}{\dn(u,m)}~. 
\ee 
These celebrated relations are known as the Landen transformation formulas. 
They have 
the special property of providing a non-trivial connection between Jacobi 
elliptic functions with unequal elliptic modulus parameters $m$ and $\tilde {m} = (1-\sqrt{1-m})^2/(1+\sqrt{1-m})^2$.

The purpose of  this paper is to give a generalization of all three Landen 
formulas. To 
describe our approach, let us focus on eq. (\ref{E3}) first. Using the identity $\dn^2(u,m)=1-m~\sn^2(u,m)$, and changing variables to $x = (1+k')u$, one can re-write the ``$\dn$" Landen formula (\ref{E3}) in the alternative form
\be\label{E4}
\dn\left[x, \left(\frac{1-k'}{1+k'}\right)^2\right]
=\frac{1}{(1+k')}~\left\{\dn \left[\frac{x}{1+k'},m \right]+dn \left[\frac{x}{1+k'}+K(m),m\right]\right\}~.
\ee
Here, the right hand side contains the sum of two terms with arguments 
separated by $K(m)$, the complete elliptic integral of the first
kind \cite{abr}. Our generalized Landen formulas will have $p$ terms on the right hand side. The results are somewhat different depending on whether $p$ is an odd or even integer. The generalization of eq. (\ref{E3}) for $p$ odd is given by 
\be\label{E5}
\dn (x,{\tilde m})
=\alpha_1 \bigg (\dn[\alpha_1 x,m]+\dn[\alpha_1 x+4K(m)/p,m]+\cdots
+\dn[\alpha_1 x+4(p-1)K(m)/p,m] \bigg )~,
\ee
where
$$
\alpha_1 = {\bigg (\dn[0,m]+\dn[4K(m)/p,m]+\cdots+\dn[4(p-1)K(m)/p,m] \bigg )}^{-1}~,
$$
\be\label{E6}
{\tilde m} = (m-2)\alpha_1^2 +2\alpha_1^3 A_1~,~~  
A_1 = \bigg (\dn^3 [0,m]+\dn^3 [4K(m)/p,m]+\cdots+\dn^3 [4(p-1)K(m)/p,m] \bigg )~.
\ee
Likewise, our generalization of eq. (\ref{E3}) for $p$ even is given by eq. (\ref{E23}). Similarly, our results for generalized ``$\cn$" Landen formulas corresponding to eq. (\ref{E2}) are eqs. (\ref{E29}) ($p$ odd) and (\ref{E33}) ($p$ even) and the generalized ``$\sn$" Landen formulas corresponding to eq. (\ref{E1}) are eqs. (\ref{E49}) ($p$ odd) and (\ref{E55}) ($p$ even). The richness of the generalized results is noteworthy - most formulas involve sums, but one [eq. (\ref{E55})] has products; most formulas have all positive signs, but one [eq. (\ref{E33})] has alternating signs; most formulas have non-trivial scalings of the argument $u$. This large variety of results is a consequence of the many different types of periodic solutions for nonlinear equations which we have previously obtained \cite{kha1,cks}.

\noindent {\bf Generalized ``dn" Landen Formulas:} Given the diversity of the generalized Landen formulas, it is necessary to establish them one at a time. To get an idea of our general approach, let us first focus on the proof of eq. (\ref{E5}).
Consider the periodic solutions of 
the static sine-Gordon field theory in one space and one time dimension,
that is, the solutions of 
\be\label{E7}
\phi_{xx} = \sin \phi~.
\ee
Note that the time dependent
solutions are easily obtained from here by Lorentz boosting. 
One of the simplest periodic solutions of the field eq. (\ref{E7}) is
given by
\be\label{E8}
\sin (\phi (x)/2) = \dn (x,{\tilde m})~.
\ee
It was shown in refs. \cite{kha1,cks} 
that a kind of superposition principle works even
for such nonlinear equations because of several highly nontrivial, new identities
satisfied by Jacobi elliptic functions \cite{kha}. In particular, it was
shown in ref. \cite{cks} that for any  
odd integer $p$, one has static periodic solutions of eq. (\ref{E7}) 
given by
\be\label{E9}
\sin (\phi (x)/2) = \alpha_1 \sum_{i=1}^{p} {\tilde d}_i~,~~p~{\rm odd}~,
\ee
where
\be\label{E10}
{\tilde d}_i \equiv \dn \left[\alpha_1 x +\frac{4(i-1)K(m)}{p},m\right]~~,
\ee
with $\alpha_1$ given by eq. (\ref{E6}).

The question one would like to address here is if solution (\ref{E9}) is completely new, or whether it can
be re-expressed in terms of simpler solutions like (\ref{E8}), but where $m$
and ${\tilde m}$ need not be the same. To that end, we note that on integrating the field eq. (\ref{E7}) once, we obtain
\be\label{E11}
\phi_x^2 = C -2\cos \phi~.
\ee
On further integration, this yields 
\be\label{E12}
\int \frac{d\phi}{\sqrt{C-2 \cos \phi}} = x+x_0~, 
\ee
where $x_0$ is a constant of integration which we put equal to zero without
loss of generality. On substituting 
\be\label{E13}
\sin (\phi /2) = \psi~, 
\ee
equation (\ref{E12}) takes the form
\be\label{E14}
\int \frac{d\psi}{\sqrt{1-\psi^2}\sqrt{\frac{C-2}{4} +\psi^2}} =x~.
\ee

Now the important point to note is that if we perform the integral for 
different values of $C$ then we will get all the solutions of eq. (\ref{E7}). Further, if
two solutions have the same value of $C$, then they must be the same. As far as
the integral (\ref{E14}) is concerned, it is easily checked that the three
simplest solutions covering the entire allowed range of
$C$ are 
\be\label{E15}
\psi = \sech ~x~, ~~C = 2~,
\ee
\be\label{E16}
\psi = \dn(x,{\tilde m})~, ~~C = 4{\tilde m} -2~,
\ee
\be\label{E17}
\psi = \cn(\frac{x}{\sqrt{{\tilde m}}},{\tilde m})~,~~C = \frac{4}{{\tilde m}}-2~, 
\ee
where $0 \le {\tilde m} \le 1$. Note that the constant $C$ has been computed
here by using eq. (\ref{E11}), which in terms of $\psi(x)$ takes the form
\be\label{E18}
C = 2 -4\psi^2 +\frac{4\psi_x^2}{1-\psi^2}~.
\ee
Thus, whereas for the solution (\ref{E16}), $C$ lies in the range $-2 \le C \le 2$, for
the solution (\ref{E17}), $C$ lies between 2 and $\infty$. Note that for 
$C < -2$, there is no real solution to eq. (\ref{E14}).

Now the strategy is clear. We will take the solution (\ref{E9}) and
compute $C$ for it and thereby try to relate it to one of the basic solutions
as given by eqs. (\ref{E15}) to (\ref{E17}). One simple way of obtaining the constant $C$ 
from eq. (\ref{E18}) is to evaluate it at a convenient value of $x$, say  $x=0$. In this way, we find 
that for the solution 
(\ref{E9}), $C$ is given by
\be\label{E19}
C = -2 + 4(m-2) \alpha_1^2 +8 \alpha_1^3 A_1~~, 
\ee
where $\alpha_1$ and $A_1$ are as given by eq. (\ref{E6}). Now, as $m \rightarrow 0$,
$\alpha_1 = 1/p,~ \dn (x,m=0) =1$ and hence $C=-2$. On the other hand, as
$m \rightarrow 1, ~K(m=1) = \infty,~ \dn (x,m=1) = \sech ~x$ and hence 
$\alpha_1 = 1$ so that $C =2$. Thus for solution (\ref{E9}), as $m$ varies in the range 
$0 \le m \le 1$, the value of $C$ varies in the range $~-2 \le C \le 2$. Hence it is clear that the solutions
(\ref{E9}) and (\ref{E16}) must be same. On equating the two $C$ values as 
given by eqs. (\ref{E16}) and (\ref{E19}), we find that the two solutions
are identical provided $m$ and ${\tilde m}$ are related by eq. (\ref{E6})
and hence the appropriate Landen transformation valid for any odd integer 
$p$ is given by eq. (\ref{E5}) with $m$ and ${\tilde m}$ being related by 
eq. (\ref{E6}). 

What about the case of even $p$? 
We have checked \cite{cks} that for this case, an exact solution is
\be\label{E20}
\sin (\phi (x)/2) = \alpha_2 \sum_{i=1}^{p} d_i~,~~p~{\rm even}
\ee
where
\be\label{E21}
d_i \equiv \dn \left[\alpha_2 x +\frac{2(i-1)K(m)}{p},m \right]~,
\ee
with $\alpha_2$ being given by 
\be\label{E22}
\alpha_2 = {\bigg (dn[0,m]+dn[2K(m)/p,m]+\cdots+dn[2(p-1)K(m)/p,m] \bigg )}^{-1}~.
\ee
Proceeding exactly as before, we find that $C$ is again in the range $-2 \le C \le 2$,  and 
hence comparing with solution (\ref{E16}) yields the Landen
transformation for even $p$:
\be\label{E23}
\dn (x,{\tilde m})
=\alpha_2 \bigg (\dn[\alpha_2 x,m]+\dn[\alpha_2 x+2K(m)/p,m]+\cdots
+\dn[\alpha_2 x+2(p-1)K(m)/p,m] \bigg )~,
\ee
where $\alpha_2$ is given by eq. (\ref{E22}) and ${\tilde m}$ is given by
\be\label{E24}
{\tilde m} = (m-2)\alpha_2^2 +2\alpha_2^3 A_2~,~  
A_2 = \bigg (\dn^3 [0,m]+\dn^3 [2K(m)/p,m]+\cdots+\dn^3 [2(p-1)K(m]/p,m) \bigg )~.
\ee
Note that when $p=2$, one recovers the Landen formula (\ref{E4}), since $\dn~[K(m),m]= k'$, and ${\tilde m}$ simplifies to $(1-k')^2/(1+k')^2$.  

\noindent {\bf Generalized ``cn" Landen Formulas:}  As shown in ref. \cite{cks}, another periodic solution of the static 
sine-Gordon eq. (\ref{E7}) is 
\be\label{E25}
\sin (\phi (x)/2) =  \alpha_3 \sum_{i=1}^{p} {\tilde c}_i~,~~p~{\rm odd}~,
\ee
where
\be\label{E26}
{\tilde c}_i \equiv 
\cn \big (\frac{\alpha_3 x}{\sqrt{m}}+\frac{4(i-1)K(m)}{p},m \big )~,
\ee
with $\alpha_3$ being given by 
\be\label{E27}
\alpha_3 = {\bigg (\cn [0,m]+\cn [4K(m)/p,m]+\cdots+\cn [4(p-1)K(m)/p,m] \bigg )}^{-1}~.
\ee
Using eq. (\ref{E18}) we can now compute the corresponding value of $C$. We obtain
\be\label{E28}
C = -2 + \frac{4 (1-2m)\alpha_3^2}{m}+8\alpha_3^3 A_3~,~~
A_3 = \bigg (\cn^3 [0,m]+\cn^3 [4K(m)/p,m]+\cdots+\cn^3 [4(p-1)K(m]/p,m) \bigg )~.
\ee
It is easily checked that since $0 \le m \le 1$, $C$ varies from 2 to
$\infty$, and hence the solutions (\ref{E17}) and (\ref{E25}) must be identical.
On equating the two values of $C$ as given by eqs. (\ref{E17}) and (\ref{E28}),
we then find that for odd $p$, the Landen transformation is given by
\be\label{E29}
\cn (x,{\tilde m})
=\alpha_3 \bigg (\cn \left[\beta x,m \right]+\cn \left[\beta x+4K(m)/p,m \right]+\cdots
+\cn \left[\beta x+4(p-1)K(m)/p,m \right] \bigg )~,
\ee
where $\alpha_3$ is given by eq. (\ref{E27}), and $\beta$ and ${\tilde m}$ are 
given by
\be\label{E30}
\beta = \frac{\alpha_3 \sqrt{{\tilde m}}}{\sqrt{m}}~,~~{\tilde m} = \frac{m}{(1-2m)\alpha_3^2 +2m\alpha_3^3 A_3}~,
\ee 
with $A_3$ being given by eq. (\ref{E28}). 

What happens if $p$ is an even integer? As shown in \cite{cks}, in that case 
another periodic solution of the static 
sine-Gordon eq. (\ref{E7}) is 
\be\label{E31}
\sin (\phi (x)/2) =  \alpha_4 \sum_{\rm i~odd}^{p} [d_i - d_{i+1}]~,~~p~{\rm even}~,
\ee
where $d_i$ is as defined by eq. (\ref{E21}) while $\alpha_4$ is given by
\be\label{E32}
\alpha_4 = {\bigg (dn[0,m]-dn[2K(m)/p,m]+\cdots-dn[2(p-1)K(m)/p,m] \bigg )}^{-1}~.
\ee
Using eq. (\ref{E18}) the value
of $C$ for this solution is easily computed and we find that $2 \le C \le 
\infty$. On comparing with solution (\ref{E17}) we find that in this case the
Landen formula is
\be\label{E33}
\cn (x,{\tilde m})
=\alpha_4 \bigg (\dn[\alpha_4 \sqrt{{\tilde m}}x ,m]-\dn[\alpha_4 \sqrt{{\tilde m}}x+2K(m)/p,m]+\cdots
-\dn[\alpha_4 \sqrt{{\tilde m}}x+2(p-1)K(m)/p,m] \bigg )~,
\ee
where $\alpha_4$ is given by eq. (\ref{E32}) while ${\tilde m}$ is given by
\be\label{E34}
{\tilde m} = \frac{1}{(m-2)\alpha_4^2 +2\alpha_4^3 A_4}~,~~ 
A_4 = \bigg (\dn^3 [0,m]-\dn^3 [2K(m)/p,m]+\cdots-\dn^3 [2(p-1)K(m)/p,m] \bigg )~.
\ee 

\noindent {\bf Generalized ``sn" Landen Formulas:} Here, we start from
the sine-Gordon field equation $(c=1)$
\be\label{E35}
\phi_{xx}- \phi_{tt}  = \sin \phi~,
\ee
and look for time-dependent solutions with velocity $v > 1$ (which are called optical soliton solutions in the context of condensed matter physics).
In terms of the variable 
\be\label{E36}
\eta \equiv \frac{x-vt}{\sqrt{v^2 -1}}~, 
\ee
the field eq. (\ref{E35}) takes the form
\be\label{E37}
\phi_{\eta \eta} = - \sin \phi~.
\ee
On integrating this equation once, we obtain
\be\label{E38}
\phi_{\eta}^2 = C +2\cos \phi~.
\ee
On integrating further, we get 
\be\label{E39}
\int \frac{d\phi}{\sqrt{C+2 \cos \phi}} = \eta + \eta_0~, 
\ee
where $\eta_0$ is a constant of integration which we put equal to zero without
loss of generality. Substituting $\sin (\phi /2) = \psi$, yields
\be\label{E41}
\int \frac{\psi}{\sqrt{1-\psi^2}\sqrt{\frac{C+2}{4} -\psi^2}} = \eta~.
\ee
If we now perform the integral for 
different values of $C$, then we get all the solutions. 
It is easily checked that the three
simplest solutions of eq. (\ref{E41}) covering the entire allowed range of
$C$ are 
\be\label{E42}
\psi = \tanh \eta~, ~~C = 2~,
\ee
\be\label{E43}
\psi = \sqrt{{\tilde m}}  
\, \sn\,(\eta,{\tilde m})~, ~~C = 4{\tilde m} -2~,
\ee
\be\label{E44}
\psi = \sn\,(\frac{\eta}{\sqrt{{\tilde m}}},{\tilde m})~,~~C = \frac{4}{{\tilde m}}-2~, 
\ee
where $0 \le {\tilde m} \le 1$. 
Note that the constant $C$ has been computed
here using eq. (\ref{E38}), which in terms of $\psi$ takes the form
\be\label{E45}
C = -2 +4\psi^2 +\frac{4\psi_{\eta}^2}{1-\psi^2}~.
\ee
Thus, for solution (\ref{E43}), $C$ is in the range $-2 \le C \le 2$, whereas for solution (\ref{E44}), $C$ lies between 2 and $\infty$. Note that for 
$C < -2$, there is no real solution to eq. (\ref{E41}).
 
Using appropriate linear superposition, it was shown in ref. 
\cite{cks} that for odd $p$ one of the solutions of eq. (\ref{E37})
is given by
\be\label{E46}
\sin (\phi (\eta)/2) =  \sqrt{m} \alpha_1 \sum_{i=1}^{p} {\tilde s}_i~,
~~p~{\rm odd}~,
\ee
where
\be\label{E47}
{\tilde s}_i \equiv \sn \left[\alpha_1 \eta+\frac{4(i-1)K(m)}{p},m\right]~,
\ee
with $\alpha_1$ being given by eq. (\ref{E6}).
Using eq. (\ref{E45}), we can now compute the corresponding value of $C$. 
We find
\be\label{E48}
C = -2 + \frac{4 m\alpha_1^2}{\alpha_3^2}~,
\ee
where $\alpha_1,\alpha_3$ are given by eqs. (\ref{E6}) and (\ref{E27}) 
respectively. 
It is easily checked that since $0 \le m \le 1$, $C$ has values between -2 and
2. Hence the solutions (\ref{E43}) and (\ref{E46}) must be identical.
On equating the two values of $C$ as given by eqs. (\ref{E43}) and (\ref{E48}),
we then find that the ``sn" Landen transformation formula for odd $p$ 
is given by
\be\label{E49}
\sn (x,{\tilde m})
= \alpha_3 \bigg (\sn[\alpha_1 x,m]+\sn[\alpha_1 x+4K(m)/p,m]+\cdots
+\sn[\alpha_1 x+4(p-1)K(m)/p,m] \bigg )~,
\ee
with ${\tilde m}$ given by
\be\label{E50}
{\tilde m} = m \frac{\alpha_1^2}{\alpha_3^2}~.
\ee 

Finally, we turn to the ``sn" Landen transformation formula for the case when $p$ is
an even integer.
One can show \cite{cks} that in this case, a solution to eq. (\ref{E37}) 
is given by
\be\label{E51}
\sin (\phi (\eta)/2) =  m^{p/2} \alpha_2 A_5 \Pi_{i=1}^{p} s_i~,
~~p~{\rm even}~,
\ee
where
\be\label{E52}
s_i \equiv \sn \left[\alpha_2 \eta+\frac{2(i-1)K(m)}{p},m\right]~,
\ee
with $\alpha_2$ being given by eq. (\ref{E22}) and $A_5$ defined by 
\be\label{E53}
A_5 = \sn [2K(m)/p,m] \, \sn [4K(m)/p,m]\cdots\sn [2(p-1)K(m)/p,m]~.
\ee
Using eq. (\ref{E45}), we can now compute the corresponding value of $C$. We obtain
\be\label{E54}
C = -2 + 4 m^{p} \alpha_2^4 A_5^4~.
\ee
It is easily checked that since $0 \le m \le 1$, the value of $C$ varies between -2 and
2 and hence the solutions (\ref{E43}) and (\ref{E51}) must be identical.
On equating the two values of $C$ as given by eqs. (\ref{E43}) and (\ref{E54}),
we then find that for even $p$, the Landen transformation formula 
is
\be\label{E55}
A_5 \alpha_2 \sn (x,{\tilde m})
= \sn[\alpha_2 x,m] \, \sn[\alpha_2 x+2K(m)/p,m]\cdots
\sn[\alpha_2 x+2(p-1)K(m)/p,m]~,
\ee
with ${\tilde m}$ given by
\be\label{E56}
{\tilde m} = m^{p} \alpha_2^4 A_5^4~.
\ee 
It is amusing to notice that as $m \rightarrow 0$, 
\be\label{E57}
A_5 (p,m=0) = \sin (\pi /p) \,  \sin(2\pi /p)\cdots\sin((p-1)\pi /p) 
= \frac{p}{2^{p-1}}~.
\ee

\noindent {\bf Transformed Modulus Parameters:} At this point, we have generalized all three of the celebrated two hundred year old 
$p=2$ Landen formulas [eqs. (\ref{E1}), (\ref{E2}), (\ref{E3})] to arbitrary values of $p$, the generalization being different depending on whether $p$ is an even or odd integer.  One 
might wonder that whereas the relation between ${\tilde m}$ and $m$ is the same
for all three ($p=2$) Landen identities, it seems to
be different for higher values of $p$. However, quite remarkably, we have established analytically 
that both for $p=3$ and for $p=4$, the relation between 
${\tilde m}$ and $m$ is in fact the same for all three generalized identities. 
For example, we find that for $p=3$, the relation between 
${\tilde m}$ and $m$ is the same for all three Landen
transformations as given by eqs. (\ref{E6}), (\ref{E30}) and (\ref{E50}). All 
expressions can be algebraically simplified and written in the common form
\be\label{E58}
{\tilde m} = m\frac{(1-q)^2}{(1+q)^2(1+2q)^2}~,
\ee
where $q \equiv \dn(2K(m)/3,m)$. Note that while deriving this result, 
use has been made of the fact that $\cn(4K(m)/3,m) = -\frac{q}{1+q}$
and that $q$ satisfies the identity
$q^4+2q^3-2(1-m)q-(1-m)=0$.
Similarly, using the relations $\dn (K(m)/2,m) = \dn(3K(m)/2,m) = t$ 
and $\dn (K(m),m) = t^2$, where
$t \equiv (1-m)^{1/4}$, it is easily
proved that for $p=4$ the relation between ${\tilde m}$ and $m$ is the 
same for all three Landen transformations as given by
eqs. (\ref{E24}), (\ref{E34}) and (\ref{E56}), i.e.
${\tilde m} = {(1-t)^4}/{(1+t)^4}~$.

What about the results for higher values of $p$? 
Although the analytic proof seems rather complicated, we have nevertheless checked numerically using the mathematical software package Maple, that the relation between 
${\tilde m}$ and $m$ is the same for all three Landen formulas with the same value of $p$. The numerical results for ${\tilde m}$ as a function of $m$ for various values of $p$ ranging from $2$ to $7$ are shown in Table 1. Note that for any fixed value of $p$, as $m$ increases from 0 to 1, ${\tilde m}$ also increases monotonically from 0 to 1. Also, for any given fixed value of $m$, ${\tilde m}$ decreases monotonically as $p$ increases.  

In all the Landen formulas (\ref{E1}), (\ref{E2}) and (\ref{E3}), the modulus parameter ${\tilde m}$ is less than the modulus parameter $m$, and this is often called an ascending Landen transformation \cite{abr}. Note that the generalizations which we have established in this paper in terms of shifts involving the period $K(m)$ on the real axis maintain the relationship ${\tilde m} < m$. We are currently working on the opposite case of descending Landen transformations and their connection with additional generalized formulas in terms of shifts involving the period $K'(m)$ on the imaginary axis.

The results of this paper clarify the relationship between the well known periodic solutions
of the various nonlinear equations and those obtained by us using the idea of judicious linear
superposition \cite{kha1,cks}. Further it provides a deep connection between the 
highly nonlinear Landen 
transformation formulas involving changes of the modulus parameter and certain linear superpositions of an arbitrary number of Jacobi elliptic functions. 

We are very grateful to the U.S. Department of Energy for providing partial support of this research under grant DOE FG02-84ER40173.

\newpage
\bigskip

\noindent {Table 1: A table of the modified modulus parameter ${\tilde m}$ in the generalized Landen transformation formulas as a function of the modulus parameter $m$ and the number of terms $p$ in the formula. Note that for odd integers $p$, the values of ${\tilde m}$ are obtained from eqs. (\ref{E6}), (\ref{E30}) or (\ref{E50}), and as mentioned in the text, they are all the same. Similarly, for even integers $p$, the values of ${\tilde m}$ are obtained from eqs. (\ref{E24}), (\ref{E34}) or (\ref{E56}), and they are also all the same. 
\sss
\sss

\oddsidemargin      -0.3in
\bigskip
\begin{tabular}{ccccccc} 
\hline
 $m$ & ${\tilde m}(p=2)$ & ${\tilde m}(p=3)$  & ${\tilde m}(p=4)$  & ${\tilde m}(p=5)$ & ${\tilde m}(p=6)$ & ${\tilde m}(p=7)$ \\
\hline
 0 & 0  & 0 & 0 & 0 & 0 & 0\\
 0.25& .5155 x$~10^{-2}$ & .9288 x$~10^{-4}$ & .1669 x$~10^{-5}$ & .3000 x$~10^{-7}$& .5392 x$~10^{-9}$& .9693 x$~10^{-11}$\\ 
 0.5& .2944 x$~10^{-1}$ & .1290 x$~10^{-2}$ & .5580 x$~10^{-4}$ & .2411 x$~10^{-5}$& .1042 x$~10^{-6}$& .4503 x$~10^{-8}$\\  
0.75& .1111 & .1005 x$~10^{-1}$ & .8666 x$~10^{-3}$ & .7438 x$~10^{-4}$& .6381 x$~10^{-5}$& .5475 x$~10^{-6}$\\  
0.9& .2699 & .4311 x$~10^{-1}$ & .6158 x$~10^{-2}$ & .8655 x$~10^{-3}$& .1213 x$~10^{-3}$& .1701 x$~10^{-4}$\\  
0.99& .6694& .2506 & .7283 x$~10^{-1}$ & .1963 x$~10^{-1}$& .5185 x$~10^{-2}$& .1362 x$~10^{-2}$\\  
0.999& .8811& .5292 & .2374 & .9312 x$~10^{-1}$& .3464 x$~10^{-1}$& .1264 x$~10^{-1}$\\  
0.9999& .9608& .7446 & .4481 & .2293& .1080 & .4891 x$~10^{-1}$\\  
0.99999 & .9874 & .8721 & .6374 & .3973 & .2239 & .1193\\  
1& 1 & 1& 1 & 1& 1& 1\\ 
\hline
\end{tabular}
\bigskip

\end{document}